%% file: photonplasma.tex
\documentclass[
    ,final            
    ,numberedheadings 
    ,sort&compress
  ]{aipproc}
\layoutstyle{6x9}
\usepackage{amssymb}
\usepackage{wrapfig}

\begin{document}

\title{Trans-Debye Scale Plasma Modeling \\ \& \\ Stochastic GRB Wakefield Plasma Processes}

\classification{98.70.Rz, 95.30.Qd, 95.75.-z, 95.30.Jx} 
\keywords      {plasma physics, kinetic modeling, GRB, wakefield processes, Monte-Carlo techniques}

\author{Jacob Trier Frederiksen}{address={Niels Bohr Institute, Juliane Maries Vej 30, 2100 K\o benhavn, Denmark}}
\author{Troels Haugb\o lle}{address={Dept. of Physics and Astronomy, \AA rhus University, \AA rhus, Denmark}}
\author{\AA ke Nordlund}{address={Niels Bohr Institute, Juliane Maries Vej 30, 2100 K\o benhavn, Denmark}}

\begin{abstract}
Modeling plasma physical processes in astrophysical context demands for both detailed kinetics and large scale development of the electromagnetic field densities.

We present a new framework for modeling plasma physics of hot tenuous plasmas by a two-split scheme, in which the large scale fields are modeled by means of a particle-in-cell (PIC) code, and in which binary collision processes and single-particle processes are modeled through a Monte-Carlo approach. Our novel simulation tool -- the \textsc{PhotonPlasma} code -- is a unique hybrid model; it combines a highly parallelized (Vlasov) particle-in-cell approach with continuous weighting of particles and a sub-Debye Monte-Carlo binary particle interaction framework.

As an illustration of the capabilities we present results from a numerical study~\cite{bib:Frederiksen2008.1} of gamma-ray burst - circumburst medium  interaction and plasma preconditioning via Compton scattering. We argue that important microphysical processes can only viably be investigated by means of such "trans-Debye scale" hybrid codes.

Our first results from 3D simulations with this new simulation tool suggest that magnetic fields and plasma filaments are created in the wakefield of prompt gamma-ray bursts. Furthermore, the photon flux density gradient impacts on particle acceleration in the burst head and wakefield.  We discuss some possible implications of the circumburst medium being preconditioned for a trailing afterglow shock front. We also discuss important improvements for future studies of GRB wakefields processes, using the \textsc{PhotonPlasma} code.
\end{abstract}

\maketitle

\input{introduction}
\input{transdebye}
\input{setup}
\input{results}
\input{discussion}

\input{acknowledgements}

\bibliographystyle{unsrt}
\bibliography{references}

\end{document}

%% file: introduction.tex
\section{Introduction}\label{ch:introduction}

\begin{figure}[ht!]
  \includegraphics[width=\textwidth]{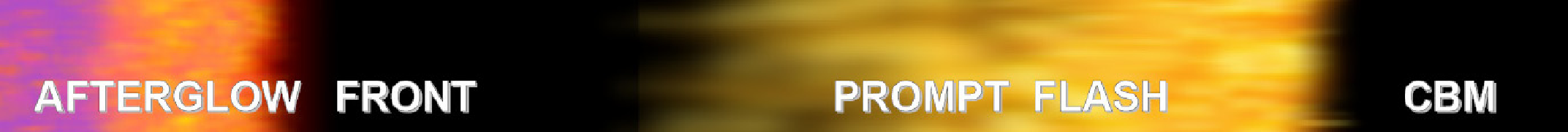}
  \caption{(Color online) \textit{Rendering -- and not to scale:} Prompt flash emission from a GRB central engine precedes a shocked ejecta, and traverses (moving right) a quiescent CBM. After the prompt gamma-ray front has passed, the CBM will relax until the ejecta shock front sweeps up the preconditioned CBM plasma. Clearly, the prompt emission front continues all the way back to the ejecta, although photon energies decrease with decreasing distance from the shocked ejecta.}
  \label{fig:promptfront}
\end{figure}

The prompt emission (PE) from a gamma-ray burst (GRB) precedes the GRB ejecta from the central engine by seconds-to-hours (depending on the phase in consideration). The PE will strongly influence the circumburst medium (CBM) before the arrival of the relativistically trailing shocked ejecta that subsequently sweeps up and shocks the CBM. We have sketched this scenario in figure~\ref{fig:promptfront}.

Lorentz factors of the GRB ejecta are generally theorized to be much higher (of order $\Gamma \approx 10^2$ - $10^3$, e.g. ~\cite{bib:PeerRyde2007,bib:Peer2008,bib:Mezsaros2006}) than that for any motion of the GRB progenitor -- and a possible accompanying wind -- itself. Both interpretation of observations  and modeling of relativistic shock physics and emission, concerned with GRB early-to-late afterglows, is complicated by its relativistic nature. The plasma microphysics of GRB afterglow shocks propagating through a CBM has been studied and extensively modeled using particle-in-cell (PIC) models in recent years \cite{bib:Spitkovsky2008,bib:Hededal2004,bib:Frederiksen2004,bib:Silva2003,bib:Gruzinov2001a}.

In contrast, modeling and observational interpretation of the PE-CBM interaction is somewhat simplified; whereas the GRB ejecta propagates at some (unknown) high Lorentz factor and involve a strong dependence on initial burst conditions, rather, the CBM is likely to be at rest relative to the progenitor from whence the burst emanates. To good approximation, this is true for the entire duration of the prompt flash. CBM rest frame events and quantities will -- after cosmological corrections -- be directly translatable to the BATSE rest frame, and since the PE is the first to reach the observer it is a strong indicator on what environment the trailing ejecta will experience. We can more directly interpret observational data and PE-CBM interaction modeling, even while relying on fewer assumptions.  Unknowns are reduced to a few: CBM density, $n_0$, prompt emission flux through a CBM surface\footnote{We simplify by assuming a slab geometry for the microphysical scenario}, and the injection photon spectrum from the prompt fireball emission ($\nu,F_\nu(t)$). 

To investigate stochastically induced wakefield effects, one may in a general way pose the following central questions concerning the PE-CBM interaction:
\begin{itemize}
    \item {{How does the PE pulse influence the CBM plasma. In particular; how does it affect particle populations and electromagnetic fields?}}
    \item {{How does the CBM plasma affect the PE gamma-ray pulse?}}
    \item {{Which physical processes are needed to adequately describe a prompt $\gamma$-ray burst signal as it traverses a circumburst plasma?}}
\end{itemize}

\noindent The rapid variability in the prompt signal has been a matter of hefty debate since the advent of GRB research in the late 1960's. The community has been arguing in favor \textit{internal} shock models on one hand, and \textit{external} shocks models on the other. The same variability has proven to be a cornerstone in establishing the relativistic expansion aspect of GRBs.

Modeling the relevant micro-physics and plasma effects is central to an understanding of whether the variability observed in the prompt phase (e.g. by BATSE) and -- more recently -- the variability observed in the early afterglow phase (SWIFT), should be attributed to an internal or external shock model, and perhaps if we need to revise these models altogether.

%% file: transdebye.tex
\section{Motivating a Choice of Model}\label{sect:transdebye}
We model, for the first time \cite{bib:Frederiksen2008.1}, the detailed interaction between a Vlasov plasma (see discussion below, and Eqn.\ref{eq:vlasovboltzmann}) and a photon burst through stochastic wakefield processes, motivated by the scenario of a prompt gamma-ray burst propagating through a quiescent circumburst medium plasma~\cite{bib:Frederiksen2008}. For this purpose, we are using a unique and highly improved particle-in-cell scheme that contains a framework to handle both Vlasov plasmas through a particle-in-cell (PIC) code described elsewhere~\cite{bib:JacobThesis,2005PhDT.........2H}, and sub-Debye sphere binary particle processes. For modeling the GRB-CBM interaction we employ only Compton scattering -- which is also described elsewhere~\cite{2005astro.ph.10292H,2005PhDT.........2H}.

\subsection{Trans-Debye Kinetic Plasma Modeling}
Photons with energies in the BATSE energy window, $10^4$ eV $\lesssim E_\gamma \lesssim 10^6$ eV, involved in scattering processes during the very first GRB-CBM interaction, will have wavelengths much smaller than: 1) the plasma interparticle distance and 2) any characteristic plasma phenomenon. Consequently, any encounter between typical plasma particles (electrons or ions) and photons can be viewed as a binary collision taking place on time and spatial scales well below the characteristic plasma scales.

Furthermore, the two plasma parameters central to a kinetic description of plasmas\footnote{Un-magnetized or weakly magnetized plasmas.} are the Debye length, $\Lambda_D$, and the electron skin-depth, $\delta_e$. Their ratio is given by $\Lambda_D~/~\delta_e \approx 1.4\cdot10^{-3}~T^{1/2}~,$ where the temperature, T, is expressed in eV \cite{bib:NRLFormulary}. This ratio is less than one for all reasonable CBM temperatures, and tends to one only for very high electron temperatures, $T \gtrsim 10^6$ eV. For the circumburst medium of GRB, it is much less than unity; for the post-flash CBM the ratio may approach unity.

\paragraph{Conventional Particle-In-Cell codes} Fundamentally PIC codes implement the Vlasov approximation; particles interact only through their self-consistently generated smoothed fields, and collisions are only of a wave-particle nature. Adding collision effects -- other than wave-single-particle effects -- ultimately leads to the Vlasov-Boltzmann formulation for plasmas
\begin{equation}\label{eq:vlasovboltzmann}
{\partial_t f} + {\bf p} \cdot {\partial_{\bf r} f} + \frac{q_s}{m_s}({\bf E} + {\bf v} \times {\bf B}) \cdot {\partial_{\bf p} f} = [{\partial_t f}]_{coll.} \stackrel{Vlasov}{\equiv}0~,
\end{equation}
where the electromagnetic forces/sources are obtained by interpolating to/from particles to the computational grid that defines the electromagnetic field. The fields are in turn integrated using Maxwell's equations, and particles are moved under the (relativistic) Lorentz force. An lower limit may be imposed on $\Lambda_D$; although PIC codes are ideally collisionless,  integration errors will introduce artificial collisions in the Vlasov-plasma, $[{\partial_t f}]_{coll.}~{\neq}~0$; the plasma heats (from phase space diffusion) until the Debye length is slightly larger than the grid cell length~\cite{bib:BirdsallLangdon}; $\Lambda_D \approx 1.5 \Delta z$. For other reasons, concerning artificial damping of collectively exicted phenomena \citep[][section 3.2ff]{bib:JacobThesis}, the Debye length should be chosen not too far away from the grid scale; a stricter limit on $\Lambda_D$ results. Consequently, we are constrained by generic particle-in-cell code properties to choose $\Delta x \approx \Lambda_D$ and all particle-particle processes such as scattering, pair production and decay are as argued all defined well below the grid scale.

\paragraph{The Random Phase Approximation} For a plasma consisting of an ensemble of point particles, $\rho(\textbf{r}) = \sum_i\delta(\textbf{r}_i-\textbf{r}_0)$, an equation of motion (e.o.m.) can be written for the electron density. Following Pines and Bohm \cite{bib:Pines1952}, we write 
\begin{equation}\label{eq:plasmaEOM}
  \stackrel{\cdot\cdot}{{\rho_k}} = 
	-\sum_i(\textbf{k}\cdot\textbf{v}_i)^2 e^{-i\textbf{k}\cdot\textbf{r}_i} 
	-\sum_{k'ij, k'\neq 0}[4\pi e^2/m(k')^2]\textbf{k} 
	\cdot 
	\textbf{k}'\left\{ exp[i(\textbf{k}'-\textbf{k})\cdot\textbf{r}_i]\right\}exp(-i\textbf{k}'\cdot\textbf{r}_i)~,
\end{equation}
where the e.o.m. is now written in terms of Fourier components. The last term can then be split into two parts, corresponding to sums over $\textbf{k}'=\textbf{k}$ and $\textbf{k}'\neq\textbf{k}$, wherein, the first part reduces to the total number of particles in the ensemble. The second part -- for all $\textbf{k}'\neq\textbf{k}$ -- reduces to $\equiv 1$ (rather, the exponential $i(\textbf{k}'-\textbf{k})\cdot\textbf{r}_i\equiv0$) under the assumption that a very large number of particles are distributed at random, spatially, and that their phases consequently average to zero. Pines and Bohm refer to this as the \textsl{random phase approximation} \cite{bib:Pines1952}. This approximation leads to a reduced expression for the e.o.m. for the density:
\begin{equation}\label{eq:reducedplasmaEOM}
  \stackrel{\cdot\cdot}{{\rho_k}} = 
	-\sum_i(\textbf{k}\cdot\textbf{v}_i)^2 e^{-i(\textbf{k}\cdot\textbf{r}_i)} 
	-(4\pi ne^2/m)\sum_ie^{-i(\textbf{k}\cdot\textbf{r}_i)}~,
\end{equation}
which for small \textbf{k} becomes the well known expression for electron plasma oscillations. For all \textbf{k} both terms in equation~\ref{eq:reducedplasmaEOM} must be taken into account and the relative strength of the two determine whether collective or single-particle effects are more important. Consequently, the random phase approximation ensures that for $k^2 \ll \Lambda_D^{-2}$ a collective description is suitable, whereas for $k^2 \gg \Lambda_D^{-2}$ the dynamics is dominated by single-particle effects. \\

\noindent It is this cross-over or split dynamics approximation that motivates our use of the hybrid scheme devised in the \textsc{PhotonPlasma} code; the PIC code handles large-scale (large compared with the Debye length) plasma effects, while the Monte-Carlo collisional framework handles collisional effects at short wavelengths. 

\subsection{Detailed Particle-Particle Processes}
Every macro-particle in PIC codes represents a large number of real physical particles, sometimes as many as $10^{30}$. Compared with previously employed PIC schemes, the \textsc{PhotonPlasma} code has been improved to include variable weighting of every single macro-particle. The charge density defined in terms of an interpolation of point particles to the field grid, for example, is then modified:
\begin{equation}
   \rho_c(\textbf{r},t) \equiv \sum_\alpha q_\alpha \sum_i     \delta({\bf r}_i - {\bf r_0},t) \\
   ~ \longrightarrow ~ 
   \rho_c(\textbf{r},t) \equiv \sum_\alpha q_\alpha \sum_i w_i \delta({\bf r}_i - {\bf r_0},t)~,
\end{equation}
with $w_i$ the weight of particles, and with $q_\alpha$ the species' charges. This seems relatively benign at first, since the gain in resolution for the macroscopic quantities is real valued but only discretely so.
The advantage becomes clear when considering scattering processes such as Compton scattering, $\textbf{p}_{e^-} + \textbf{p}_{\gamma} \rightarrow \textbf{p}_{e^-} + \textbf{p}_{\gamma'}$, used in the GRB wakefield experiments presented in the following sections. With variable weighting, the \textsl{physical} number densities of the plasma constituents are rather now given in some volume, e.g. a cell volume, by 
\begin{equation}
    n_e = \sum^{N_{e,cell}}_{i=1} w_i^{(e)} ~ ~, ~ ~ n_\gamma = \sum^{N_{\gamma,cell}}_{j=1} w_j^{(\gamma)} ~,
\end{equation}
for electrons and photons respectively. We can now define the physical Compton scattering rate in that same volume as
\begin{equation}
\frac{\partial N_{e,\gamma}}{\partial t} 
  = \int_{V} \frac{\partial n_{e,\gamma}}{\partial t} c dV 
  = \int_{V} \sum^{N_{e,cell}}_{i=1} \sum^{N_{\gamma,cell}}_{j=1} w_i^{(e)} w_j^{(\gamma)} \overline{\sigma}_C ({\bf p}^{(e)}_i,{\bf p}^{(\gamma)}_j) c dV
\end{equation}
with $\overline{\sigma}_C$ the Compton scattering cross section. We have assumed here that the number of scatterings is proportional to the product number densities of the interacting species. This leads to a change in number density of the original interacting particles, $\Delta n_{e,\gamma} \propto n_e n_\gamma \overline{\sigma}_C ({\bf p}_e,{\bf p}_\gamma) c \Delta t$, where resolution gain is now real valued \emph{and} continuous.
The number of (Compton) scatterings is then given as a \textsl{continuous} real number quantity, $\Delta n_{e,\gamma}$ on the cell volume, even if the particles are discrete entities. 

Scattering is achieved by splitting the macro-particle pair weights into a total of four new particles, two of which carry the new scattered momenta and energies as determined by the microscopic physical cross section ($\overline{\sigma}_C ({\bf p}_e,{\bf p}_\gamma)$). This is manifested through $w^{(e,\gamma)}_{old} = w^{(e,\gamma)}_{new} + w^{(e,\gamma)}_{scatt}~,$ for the electron and photons both.  An important fact about this scattering procedure is that we retain full resolution in change of momenta, energies, weights, charges decay rates etc.  For example, consider the scattering in a volume of $10^{30}$ photons and electrons. Effectively, for a scattering fraction of $10^{-6}$, $10^{24}$ physical particles are scattered away from a 'mother' particle of size $10^{30}$. This illustrates the flexibility and the fundamentally continuous nature of the trans-Debye kinetic approach. \\

\noindent For the modeling of GRB prompt emission interacting with the tenuous and relatively cold circumburst medium, the considerations summarized in this section motivated the development and deployment of the \textsc{PhotonPlasma} code.

%% file: setup.tex
\section{Setup -- Stochastic Wakefield Simulation}
\begin{figure}[h!]
    \centering
    \includegraphics[width=\textwidth,height=.2\textwidth]{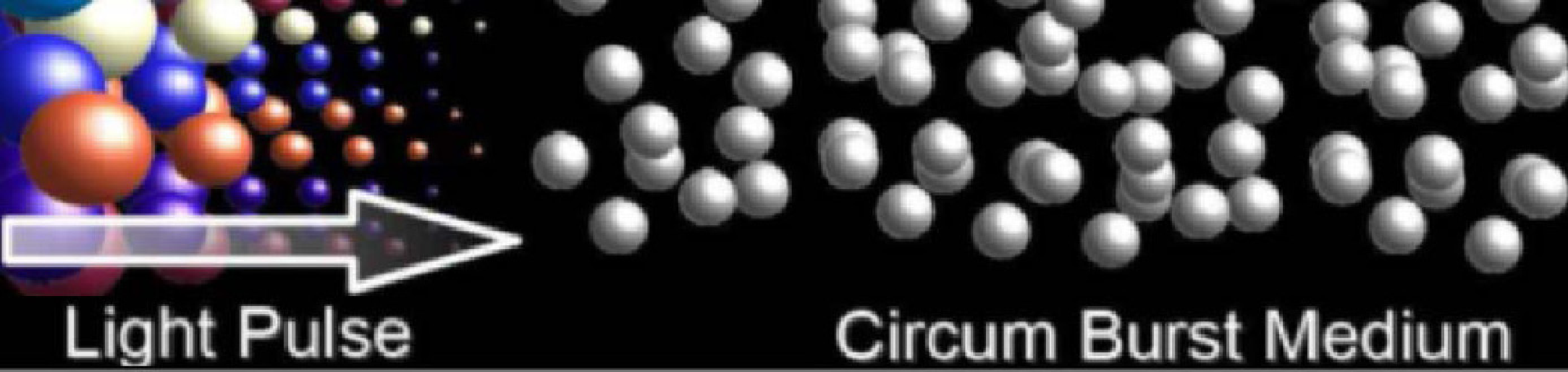}
    \caption{(Color online) Rendering of the prompt GRB photons (colored particles) traversing the quiescent CBM (grey particles) in our model. Photons are added to the computational domain on the left boundary. As the photon number flux density increases, the \textit{weights} (size of particles) -- rather than the number -- of the photons vary in a continuous way. Color designates that photons further carry different energies (through $\nu$) as well. This detailed balance approach saves computational effort and increases speed. The microphysics is unchanged (section~\ref{sect:transdebye}).}
\label{fig:lightpulse}
\end{figure}

A synthetic thermal GRB is delivered to the quiescent CBM plasma by adding photons (particles) to the computational domain on the left volume boundary, according to a prescribed light curve and spectrum -- this is sketched in figure~\ref{fig:lightpulse}.

\begin{figure}[h!]
		\includegraphics[width=\textwidth,height=.25\textwidth]{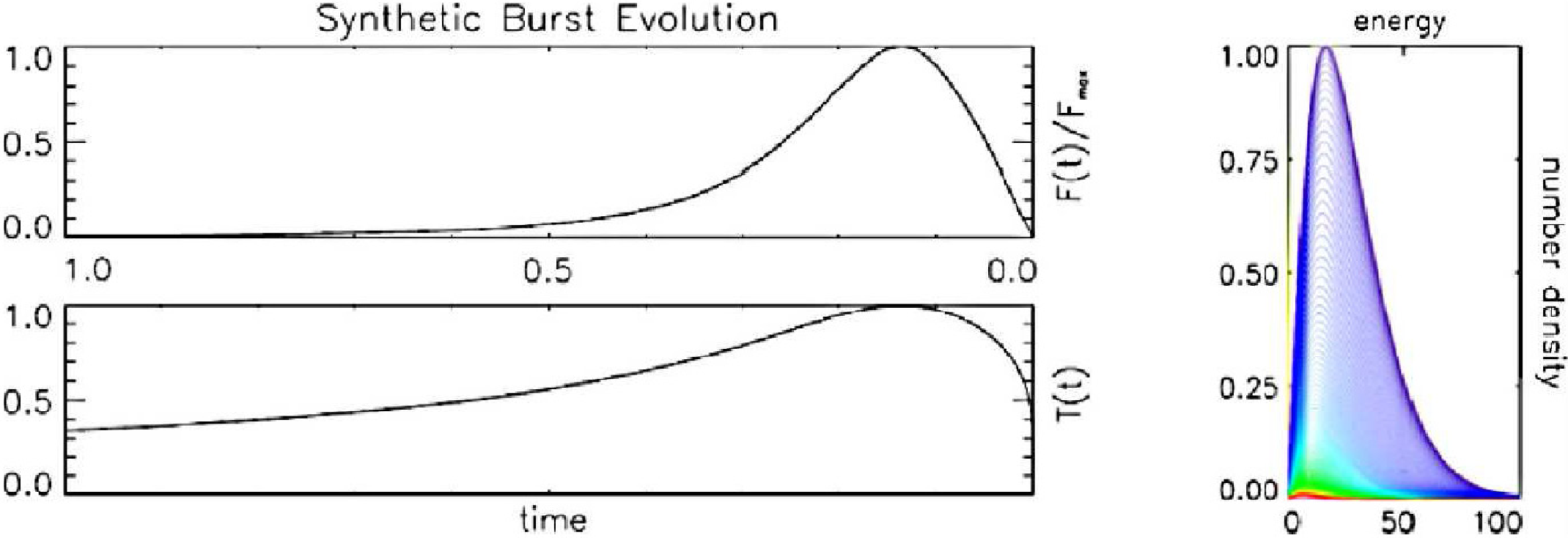}
	\caption{(Color online) \textit{Left:} synthetic GRB fast-rise-exponential-decay (FRED) light curve (top), and the temperature evolution (bottom) of the PE, assuming perfect blackbody radiation. Time, flux and temperature are normalized to the total burst duration, peak flux and temperature at peak flux -- respectively. \textit{Right:} the spectral photon radiance of the thermal burst for various times. Photon energy and number density is scaled in \textsl{per cent} of BATSE energy window and peak radiance -- respectively. Color of individual curves designates linear time from burst onset.}
	\label{fig:fig3}
\end{figure}

\paragraph{Gamma-Ray Burst Synthesis} The light curve for the synthetic gamma-ray burst is inspired by observational data fitted to a FRED function \cite{bib:Ryde2003,bib:Kocevski2003} with parameters chosen for BATSE trigger 3891 (GRB951102), $r=1.26$ and $d=2.67$. We have set the duration of the photon burst to $\tau_{GRB} = 200 \omega_{pe}^{-1}$, which corresponds to about 1\% of $T_{50}$ for trigger 3891. Assuming a pure blackbody spectrum, the accompanying temperature curve is shown in the lower left panel of figure~\ref{fig:fig3}.

A considerable sample from the BATSE catalogue may be modeled with better statistic assuming that the prompt emission spectrum is composed of a power law component, overlayed with a blackbody spectral component (see e.g. \cite{bib:Ryde2004}). Temperatures are found to be of order $T_{BB} \sim 50 \textnormal{keV}$. This is well below pair threshold and -- in the first approximation -- we may neglect pair production as an important effect during late time evolution of the prompt burst\footnote{Note, however, that Svensson~\cite{bib:Svensson1984} found that even for these moderate temperatures, pair production could be significant in the high-energy tail of the particle distribution -- see also \cite{bib:Ghisellini1999}.}. Photon energies for the thermal evolution are normalized to the BATSE observational window, $E_\gamma \in [33.6$keV$,3.36$MeV$]$ which is corrected for a luminosity-lag estimated redshift of $z\sim1.68$~\cite{bib:Norris2001,bib:Ryde2003}.

Temporal evolution of the GRB is achieved through the assignment of variable weighting to a \textsl{constant} number flux of computational macro-particle photons (see also section \ref{sect:transdebye}), this corresponding to a \textsl{varying} number flux of physical photons, in accordance with the prescribed light curve (particle weights) and spectrum (particle energies). Photon energies are sampled from MC integration of the (time-dependent) Planckian; we obtain a comparatively high accuracy at low computational cost. The photon spectral radiance (spectral photon number density) is given by $$n_\gamma(\nu,t) \propto \frac{F(\nu,t)}{\nu} \propto \sum^{N_{\gamma}(\nu,t)}_{i=1} w_i(\nu,t) ~ \Rightarrow ~ n_\gamma(\nu,t) \propto \sum^{N_{\gamma}(\nu)}_{i=1} w_i(t)~.$$ 

Through split assignment we impart time-variability to the weights, $w_i(\nu,t) \rightarrow w_i(t)$, and frequency variability to individual particles, $N_\gamma(\nu,t) \rightarrow N_\gamma(\nu)$. This yields a high degree of flexibility compared to conventional PIC codes as explained in detail in section~\ref{sect:transdebye}. The macro-particle photon number density is set constant at 40 particles/cell for the entire duration of the burst.

\paragraph{Circumburst Medium Plasma Model and Boundary Conditions} 
We assume the CBM to be a quasi-neutral thermal hydrogen plasma (ions and electrons) of constant density with moderate temperature, $v_{th,e}\approx0.1c$, and slightly reduced mass ratio of the plasma constituents of $m_i/m_e = 256$.  The plasma is initially thermal and field free (except for thermal fluctuations). No kinetic photons are present in the computational volume. Boundaries are periodic in the (X-Y-) plane transverse to the GRB pulse propagation.  A grid resolution of 100$\times$20$\times$4000 translates to \{$L_x, L_y, L_z$\} $\approx$ \{13$\delta_e\times$2$\delta_e\times$500$\delta_e$\} for our choices of length units and plasma density. Initially, the number of plasma particles is 40/species/cell, but the total number of particles increases as scattering splits the interacting particle pairs during the run. Using a relatively flat volume aspect ratio is a compromise to ensure skin-depth resolution at the lowest possible grid size to give any 3D structure a chance of growing marginally without spending excess computing time. The moderate number of particles/cell has minor impact on resolution issues as discussed in section~\ref{sect:transdebye}, since scattering adds particles and conserves the momentum locally everywhere, and globally conserves energy to better than $1\%$. The burst duration in light travel length is $L_{GRB} \approx 64\delta_e$. The lower/upper Z-boundary (GRB influx/outflux boundary) specularly reflects/absorbs all plasma, but both ends allow for photons to exit. Electromagnetic fields are damped in a thin layer, $\Delta L_z \sim 0.1\%L_z$ on both upper and lower Z-boundaries and are periodic in the tranverse direction (XY), to avoid any spurious EM wave growth.

%% file: results.tex
\section{Results -- Stochastic GRB Wakefield Effects}
\begin{figure}[h!]
		\includegraphics[width=\textwidth]{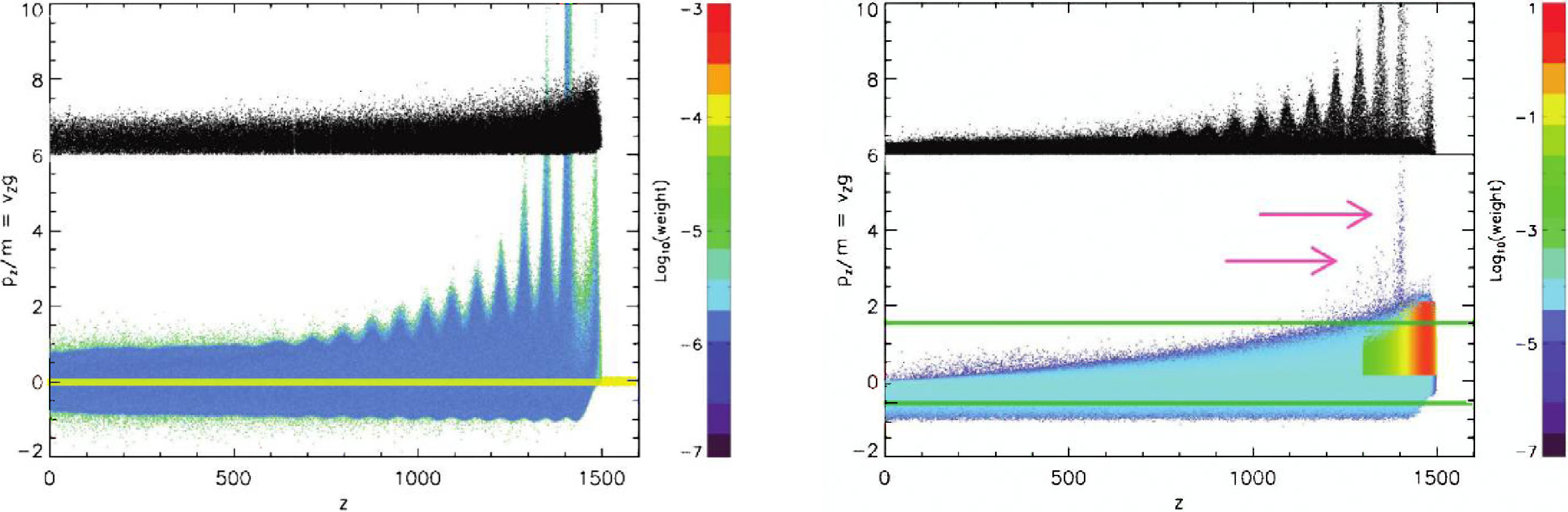}
  \caption{(Color online) \textit{Left -- electron scatter plot}:  Color designates weight in log-scale -- blue particles have weights $w_{blue} \sim 10^{-5}w_0$, where $w_0$ is the initial weight. The yellow line coincides with zero bulk z-momentum, $p_{z,\gamma}=0$. For clarity photons (black) are offset vertically by $p_{z,\gamma}=+6$. \textit{Right -- photon scatter plot}: The initial burst of photons (in color -- moving right) is visible from $t\sim1500$ to $t\sim1300$. Pink arrows: local photon upscattering. Green lines: energy threshold for $\gamma+\gamma \rightarrow e^++e^-$ for photons pairwise oppositely outside (away from $p_z=0$) these lines. Electrons (black) are offset by $p_{z,e}=+6$. Here, $p_{z,\gamma} \equiv [h\nu]_z/m_ec^2$.}
	\label{fig:fig2}
\end{figure}

\begin{figure}[!h]
		\includegraphics[width=\textwidth]{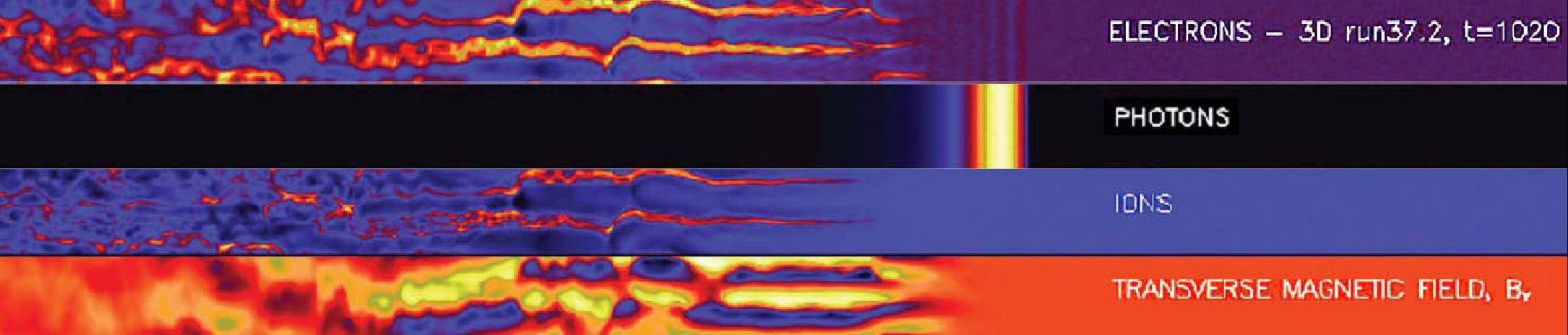}
  \caption{(Color online) Current structures for electrons (panel 1) and ions (panel 3), the photon pulse (panel 2), and (transverse) magnetic field (panel 4) in our 3D wakefield simulation, for $t=1020\omega_{pe}^{-1}\approx 10^4\Delta t$. The GRB photon pulse has light-length $L_c \equiv c/\tau_{GRB} = 200 \approx 64\delta_e$. The computational volume has dimensions \{$L_x, L_y, L_z$\} = \{40, 8, 1600\} or, in electron skin-depths, \{$L_x, L_y, L_z$\} $\approx$ \{12$\delta_e$, 2$\delta_e$, 500$\delta_e$\}.}
	\label{fig:fig6}
\end{figure}

Here we present results only from the largest run in our wakefield simulation batch, with parameters as described in the previous section. For a more complete treatment of various effects of changes to the setup, please refer to \cite{bib:Frederiksen2008} and \cite{bib:Frederiksen2008.1} where also a number of tests and details are given. Robustness of the central results presented here was tested through a consistent scan of parameter space. We concentrate here on the plasma-particle heating/acceleration aspects and electromagnetic field growth. We shall distinguish between 'plasma particles' (electron macro-particles) and 'photons' (photonic macro-particles), see also section~\ref{sect:transdebye}. Figures~\ref{fig:fig2} and \ref{fig:fig6} compactly summarizes the results of stochastically forcing the quiescent thermal plasma -- the CBM -- by a strong thermal photonic pulse -- the GRB. 

\paragraph{Wakefield -- Particle Effects}

As the photon pulse interacts with the plasma through Compton scattering, it excites waves in the plasma. Although excited by single macro-particle scatterings in a highly optically thin system ($\tau \ll 1$; the integrated weight of photons scattered at least once per wavelength is less than $10^{-3}$), these waves are the collective reaction of the plasma to the photon stochastic forcing. We emphasize the random nature of the scattering. 

A strong acceleration of the particles in the wakefield of the photon pulse is seen in Figure~\ref{fig:fig2} -- right panel, where plasma particles are plotted in color. To sustain quasi-neutrality in the wakefield plasma, a subset of the plasma particles (taken from those not scattered) are accelerated backwards, $p_z < 0$, in the wakefield. Their response is collective as no preference is given to any one particle in the Vlasov-formulation; all processes are described by characteristic scales, $k_C$, above the Debye-length, $k_C > \Lambda_D^{-1}$.

Plasma structures excited by the photon pulse are almost co-moving with the pulse, lagging gradually behind the pulse front due to differences in propagation speed of the photons and Langmuir waves. The structures are therefore dragged out behind the first photon-plasma encounter, all while growing in strength (height), and they increasingly interact with the photon pulse; a feedback loop to the photon pulse produces the photon energy spikes shown in the right panel of Figure~\ref{fig:fig2}. 

\paragraph{Wakefield -- Electromagnetic Field Effects}
A counter-streaming plasma results from the anisotropic nature of the Compton interaction as seen in the left panel of Figure~\ref{fig:fig2}, where the momentum is centered on $|\textbf{p}_z| = 0$, and we may expect the development of relativistic (or mildly so) streaming instabilities, such as filamentation and the two-stream instability (see e.g. \cite{bib:Dieckmann2006,bib:Bret2005}). Figure~\ref{fig:fig6} displays a cross section through the computational domain, revealing the production of current filaments (filamentation) and an associated magnetic field, with an energy density equal to a few \textsl{per cent} of equipartition, $\varepsilon_B \approx 5\%$, with the total kinetic energy transfer fed to the plasma. While ions play the role of maintaining the filaments, the electrons produce the current necessary for the magnetic field to exist. We have tested the robustness of the field generation for various burst durations and luminosities, and found B-field generation in all cases~\cite{bib:Frederiksen2008}. 

\begin{wrapfigure}{r}{2.5in}
\begin{minipage}[f]{2.5in}
\begin{center}
	\includegraphics[width=2.25in]{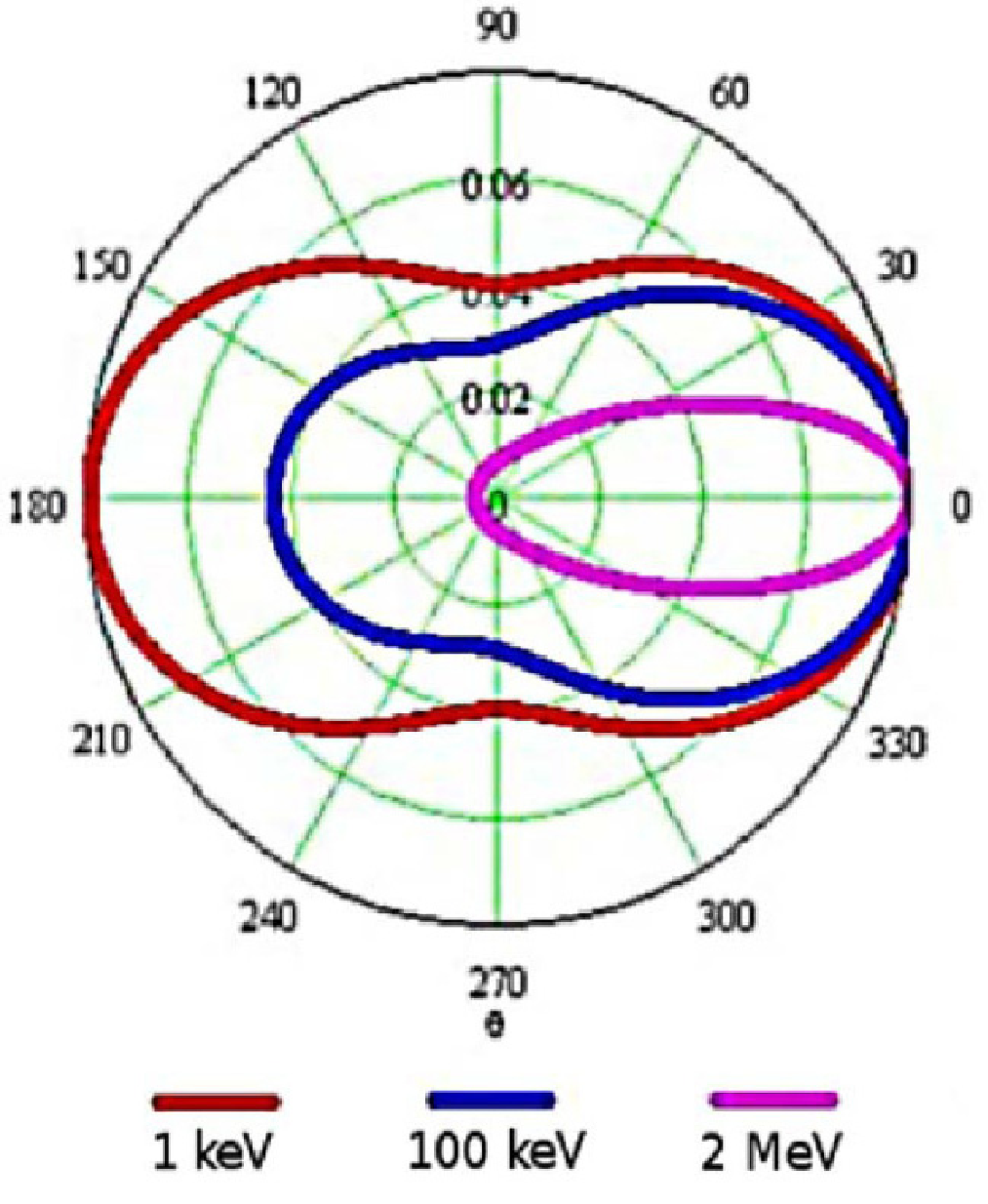}
\label{fig:comptoncross}
\end{center}
\end{minipage}
\end{wrapfigure}

\paragraph{A Photon Anisotropy Effect}
The energy exchange in Compton scattering is $$ \epsilon_2 = \frac{\epsilon_1}{1 - \frac{\epsilon_1}{m_ec^2}(1 - \cos \theta)}~.$$ Forwardly scattered photons exchange rather small amounts of energy with the (rest) electron whereas reversely scattered photons may exchange large amounts of energy. The figure inset shows differential Compton cross section \cite{bib:UniverseReview} for energies spanning the BATSE observational energy window. High energy photons (electron rest frame) will scatter mainly forward and low energy photons mainly either backwards or forwards.

We may now explain the apparent stability of the CBM counter-streaming plasma set up by the GRB pulse: photon scattering on electrons moving forward (pulse-parallel) will scatter severely and impart energy to the electron so, as to accelerate it even more (CBM rest frame). For a photon scattering on an electron moving backward (pulse-anti-parallel), however, the momentum and energy exchange in the scattering would be such that no significant acceleration of the electron would result. In a second scattering generation the process would be repeated (with slightly different energies, angles etc.). Consequently, counter-streaming is maintained -- at least for the duration of the prompt burst; the CBM slowly 'eats' momentum anisotropy from the GRB pulse.

Whether the magnetic fields generated by the passing GRB photon burst are quasi-static to the extent that a trailing shocked ejecta could be influenced by the plasma and field structures produced, has yet to be determined.  Field structures are sustained, we speculate,  at least as long as significant photon momentum anisotropy is present.

%% file: discussion.tex
\section{Discussion and Conclusions}
In conclusion, we have presented a new split dynamics framework -- the \textsc{PhotonPlasma} code -- for modeling the plasma physics of hot tenuous plasmas. Large scale fields are modeled by means of a PIC code. Collisions and other sub-Debye processes, are modeled through a detailed Monte-Carlo approach.

Our first wakefield simulation results using the code suggest that magnetic fields and plasma filaments are created in the wakefield of a prompt GRB. Particles are heated and accelerated and photons scatter into a distribution capable of photon-photon production.

We identify two questions to be answered in the future, before significant progress can be made on GRB-CBM modeling and relativistic afterglow shock modeling:
\begin{enumerate}
	\item Can the plasma in the GRB wakefield undergo magnetic hysteresis -- i.e. can it become magnetized, and stay magnetized, on timescales comparable to the time gap between the afterglow shock front and the GRB wakefield?
	\item Which binary particle processes, other than Compton scattering, are inclusions imperative to more further investigation of the effects found in the present work?
\end{enumerate}

The first of these questions is likely to require the extension of the modeling framework to include the transition from Vlasov-kinetic (PIC) codes to a coarser approximation in the plasma physics formulation hierarchy (i.e. Vlasov-fluids, Hall-MHD, \ldots).

Regarding the second question; from figure~\ref{fig:fig2}, right panel, we see that counter-streaming pairs of photons (above and below the green lines) are above pair production threshold. Such photon pairs are not observed in similar runs without the plasma effects included. We therefore argue that long-time development of the GRB-CBM system must include both plasma effects and photon-photon pair production.

The energetic photons in the spikes of fig.~\ref{fig:fig2} (right panel) will, if they survive traversing the CBM, be observed as non-thermal emission. Although the color of these spikes indicate weights only of order $10^{-4}$ and less, it should be noted that these spiky structures accumulate with time. They may eventually after passing a much longer distance lead to significant influence on the GRB spectrum away from its initial profile.

Even if the interaction seemingly is locally unimportant for the burst photon front, such plasma and pair production effects could eventually affect -- in an accumulative way -- the final resulting observed GRB prompt burst.

%% file: acknowledgements.tex
\begin{theacknowledgments}
We thank Juri Poutanen for funding the workshop\footnote{Thinkshop 2005, Stockholm 2005; http://comp.astro.ku.dk/Twiki/view/NumPlasma/ThinkShop2005} that initialized the development of the \textsc{PhotonPlasma} code. Further, we thank Boris Stern for ample help with Compton scattering Monte-Carlo techniques. Collaboration with Dr. Christian Hededal on the simulation tools is acknowledged. JTF acknowledges support provided by the Swedish Research Council, and ÅN and JTF acknowledges support from the Danish Natural Science Research Council. Computational resources and support were provided by the Danish Center for Scientific Computing (DCSC).
\end{theacknowledgments}